\documentstyle[colordvi,prl,twocolumn,aps,epsfig,epsf]{revtex4b1}
\begin{document}

\draft

\title{Ergodicity of Random--Matrix Theories: The Unitary Case}
\author{Z. Pluha\v r$^1$ and H. A. Weidenm\"uller$^2$}
\address{$^1$Charles University, Prague, Czech Republic}
\address{$^2$Max-Planck-Institut f\"ur Kernphysik, D-69029
Heidelberg, Germany}


\maketitle

\vspace*{-0.5cm}

\begin{abstract}
We prove ergodicity of unitary random-matrix theories by showing that 
the autocorrelation function with respect to energy or magnetic field
strength of any observable vanishes asymptotically. We do so using
Efetov's supersymmetry method, a polar decomposition of the saddle--point
manifold, and an asymptotic evaluation of the boundary terms generated
in this fashion.
\end{abstract}


Predictions of random--matrix theories (RMT) are obtained by averaging
over the ensemble of random matrices. To test such a prediction, one
compares it with data obtained from a single system (rather than from
an ensemble of systems governed by different Hamiltonians). For the
data set, the ensemble average is replaced by the running average
(i.e., the average over energy $E$, magnetic field strength $B$, or 
another variable). If the two averages are, in a physically plausible
limit, almost always equal irrespective of the observable at hand, RMT
is said to be ergodic.

It is generally believed that RMT is ergodic. However, proofs have
been given only for spectral fluctuations and some other special
observables, see the review \cite{guh98}. In this Letter, we report on
the first general proof of ergodicity for the Gaussian Unitary
Ensemble (GUE), the ensemble of Hermitean matrices $H_{\mu \nu}$ of
dimension $N$ where $N \rightarrow \infty$. The matrix elements are
Gaussian random variables with zero mean and a second moment given by
$\overline{H_{\mu \nu}H_{\mu' \nu'}} = (\lambda^2/N) \delta_{\mu \nu'}
\delta_{\mu' \nu}$. Our proof is not restricted to this ensemble,
however, and applies equally to all other non--Gaussian unitary
ensembles specified in the paragraph below Eq.~(\ref{eq1}). This is
because all these ensembles are known \cite{hac95} to possess
identical local fluctuation properties. It is these fluctuation
properties which are used in the proof. In the framework of the GUE,
we consider the two most important cases: The running average extends
either over energy or over magnetic field strength. 

The proof is based on the use of Efetov's supersymmetric generating
functional \cite{efe83,ver85} and on a polar decomposition of the
saddle--point manifold. The latter yields a series of boundary or
Efetov--Wegner terms which we generate using Refs.~\cite{gos98,plu99}.
The restriction to the unitary case is for technical reasons only. 
We feel certain that the method of proof works equally for the
technically more demanding orthogonal and symplectic ensembles. Lack
of space forces us to give the essential steps only. Details may be
found in Ref.~\cite{plu00}.

In the standard approach to the problem \cite{pan79} one considers an
arbitrary observable $F$. To be specific, we assume that $F$ is a
function of energy $E$, so that $F = F(E,H)$ where $H$ denotes the
Hamiltonian. We return to the case of the magnetic field strength
at the end of the paper. We denote the ensemble average (the running
average) by an overbar (an angular bracket, respectively). Explicitly,
$ \overline{F(E)} = \int \mbox{d}\mu(H) F(E,H) $ where
$\mbox{d}\mu(H)$ denotes the GUE measure for integration over $H$
(including the Gaussian weight factor), and $ \langle F(E) \rangle =
\Delta^{-1} \int_{E - \Delta /2}^{E + \Delta /2} \mbox{d} E' F(E',H)
$ where $\Delta$ denotes the averaging interval, and where $H$ is an
arbitrarily chosen member of the random--matrix ensemble. To define
ergodicity, one asks whether
\begin{equation}
\label{erg}
\overline{  \Bigl ( \overline{F(E)} -
  \langle F(E) \rangle \Bigr )^2 } \rightarrow 0 $ for all $F(E,H) \ .
\end{equation}
The arrow denotes a limit defined below. If this condition is met, the
difference between the two averages tends almost always to zero, and
the ensemble is said to be ergodic.

The GUE spectrum is confined to the interval $-2\lambda \leq E \leq 
2\lambda$, so that the average level spacing $d \sim \lambda/N$. In
the ergodicity test (\ref{erg}), one considers the limit where the
running average (RA) is performed {\it after} the limit $N \rightarrow
\infty$ has been taken. The RA extends over $K$ levels where $K
\rightarrow \infty$ and, thus, over an energy interval of length
$\Delta = K d$ which, for any finite $K$, is infinitesimally small in
comparison with the range $4 \lambda$ of the spectrum. The ensemble
average $\overline{F(E)}$ depends on $E$ only via the mean level 
density $\rho(E)$ and is, on the scale $K d$, therefore constant
(local stationarity). It follows that $ \overline{ \langle F(E) \rangle
  } = \overline{F(E)} $. This fact allows us to cast the ergodicity
condition in the form $\overline{ \langle F(E) \rangle^2 } - \Bigl(
\overline{F(E)} \Bigr)^2 \rightarrow 0 $. The term $\overline{ \langle
  F(E) \rangle^2 }$ can be written as a double RA taken over the
function $\overline{F(E_1)F(E_2)} = \overline{F(E_1,H)F(E_2,H)}$. This
function is symmetric in $E_1, E_2$, and, for $E_1,E_2 \subset
\Delta$, depends only on $|E_1 - E_2|$. With the angular brackets
denoting now this double RA, the ergodicity condition reduces to
\begin{equation}
\label{erg2}
\Bigl \langle \overline{F(E_1)F(E_2)} - \Bigl( \overline{F(E)} 
\Bigr)^2 \Bigr \rangle \rightarrow 0 \ .
\end{equation}
Application of Slutsky's theorem~\cite{pan79,yag62} shows that
ergodicity holds provided the autocorrelation function $C_{\cal E} =
\overline{F(E_1)F(E_2)} - \overline{F(E_1)} \ \overline{F(E_2)}$ 
vanishes for large $|E_1 - E_2|/d$. This is what we are going to show.

In general, $\overline{F(E,H)}$ may itself be a $k$--fold correlation
function. Then, $F(E,H)$ will primarily not depend on $E$ but on the
$k$ energy arguments $\epsilon_1, \ldots, \epsilon_{k}$ where $k$ is a
positive integer. To implement the ergodicity limit, we write
$\epsilon_j = E + \omega_j$ with $\sum_j \omega_j = 0$ and $E = (1/k)
\sum_j \epsilon_j$. The $\omega_j$'s are defined on the scale $d$ and
are held fixed, and we consider $F$ as a function of $E$ only. Using
standard notation and procedure \cite{ver85}, we can cast $\overline{
  F(E_1)F(E_2)}$ in the form
\begin{equation}
\label{eq1}
\overline{F(E_1)F(E_2)} = \int \mbox{d} \mu (T) \exp \bigl ( i \pi 
[ \varepsilon / d ] {\rm trg} ( Q \tau_3 ) \bigr ) S( Q ) \ .
\end{equation}
Here, $Q = T^{-1} L T$ is a graded matrix of dimension $8k$,
$\mbox{d}\mu (T)$ is the invariant measure for integration over the
matrices $T$ belonging to the coset
$\mbox{U}(2k,2k/4k)/[\mbox{U}(2k/2k)]^2$, and ${\rm trg}$ denotes the
graded trace. The $8k$ matrix indices are labelled $p r j \alpha$
which follow in lexicographical order. Here, $p = 1,2$ distinguishes
the retarded and advanced form of the Green functions, $r = 1,2$ the
energy arguments $E_1$ and $E_2$, $j = 1,\ldots,k$, and $\alpha$ is
the supersymmetry index which assumes the valu
es $b$ for Bosons and
$f$ for Fermions. The matrices $L$ and $\tau_3$ are diagonal matrices
of dimension $8k$ with entries $\pm 1$. In the matrix $L$ ($\tau_3$),
the two signs distinguish retarded and advanced Green functions (the
two observables $F(E_1,H)$ and $F(E_2,H)$, respectively). We have
defined $\varepsilon = (E_1 - E_2)/2$. Without loss of generality, we
take $\varepsilon > 0$ in the sequel. The explicit form of $S$ depends
on the form of the observable $F(E,H)$ under consideration and is
immaterial for what follows. Suffice it to say that $S$ does not
depend upon $\varepsilon$, and that it contains a factor which depends
exponentially on all the energy arguments $\omega_j$ with $j =
1,\ldots,k$ and a second factor which contains the source terms and,
possibly, the coupling to open channels.

Given Eq.~(\ref{eq1}), the proof of ergodicity is a non--trivial and
exact technical exercise which is sketched in the remainder of the
paper. The limits of validity of the proof are those of
Eq.~(\ref{eq1}): We only admit ensembles of unitary matrices for which
(i) the spectrum is confined to a simply connected compact domain, and
for which (ii) the fluctuation properties are translationally
invariant (locally stationary) in energy. Counterexamples to (i, ii)
are soft confining potentials~\cite{mu93} which typically have
non--stationary fluctuation properties and for which another proof of
ergodicity would be required. Counterexamples to (ii) are chiral
random--matrix models where the point $E = 0$ plays a distinct
role. It is known that ergodicity does not hold near this point.
(iii) Eq.~(\ref{eq1}) is restricted to the interior domain of the
spectrum. Exclusion of the end points is due to a technical limitation
of the method~\cite{efe83,ver85}. These points are of little interest
for most applications of random--matrix theory where the energies
$E_1, E_2$ are chosen to lie at or near the center of the spectrum.

Eq.~(\ref{eq1}) looks like a Fourier integral. Given sufficient
regularity of $S$ near ${\rm trg} (Q \tau_3 ) = 0$, we expect
$\overline{F(E_1)F(E_2)}$ to vanish asymptotically for large
$\varepsilon/d$, except for a term which cancels $\overline{F(E_1)} \
\overline{F(E_2)}$. Our task consists in transforming this qualitative
expectation into a proof. To this end we adopt the procedure of
Ref.~\cite{alt93} and simplify the term ${\rm trg} ( Q \tau_3 )$ in
Eq.~(\ref{eq1}) in two steps. First, we write $T$ as the product $T =
T_C T_D$ where $T_C $ ($T_D$) does not (does) commute with $\tau_3$.
With $Q_C = T_C^{-1} L T_C$, this yields ${\rm trg} ( Q \tau_3 ) =
{\rm trg} ( Q_C \tau_3 )$. Technically, the matrices $T_C$ ($T_D$) are
obtained by exponentiating the coset generators which anticommute
(commute) with $\tau_3$. The result can be expressed in terms of the
$4k \times 4k$ graded matrices
\begin{equation}
T_{q} 
= \left(
  \begin{array}{cc}
  t_{q}^{11} & t_{q}^{12}\\[6pt]
  t_{q}^{21} & t_{q}^{22}
  \end{array} 
  \right)
\label{Tq-t}
\end{equation}
with $q=1,...,4$. These matrices belong to the coset $U(k,k/2k) /
[U(k/k)]^2$. The matrix elements of $t_q^{12}$ and their conjugates in
$t_q^{21}$ represent the ``Cartesian'' coordinates of $T_q$. As usual,
$t_q^{12}$ and $t_q^{21}$ are symmetry--related. Moreover, we have
$t_q^{11} = (1 + t_q^{12} t_q^{21})^{1/2}$ and $t_q^{22} = (1 +
t_q^{21} t_q^{12})^{1/2}$. The matrices $T_D$ ($T_C$) are given in
terms of $T_1, T_2$ ($T_3, T_4$, respectively). Interest focuses on
$T_C$ which has the form
\begin{equation}
\Red T_{C}
= \left(
  \begin{array}{cccc}
  t_{3}^{11}  &  0  &  0   & t_{3}^{12}   \\
  0  &  t_{4}^{11}  &  t_{4}^{12}  &  0   \\
  0  &  t_{4}^{21}  &  t_{4}^{22}  &  0   \\
  t_{3}^{21}  &  0  &  0   & t_{3}^{22}
  \end{array}
  \right)  \;.
\label{TDTC-t}
\end{equation}
A second simplification of ${\rm trg} ( Q \tau_3 )$ arises when,
following Efetov \cite{efe83}, we express $T_s$ for $s = 3,4$ in
``polar coordinates'' by diagonalizing $t_s^{12}$ and $t_s^{21}$. We
write $t_s^{12} = u_s^1 \lambda_s (u_s^2)^{-1}$ and $t_s^{21} = u_s^2
\overline{\lambda}_s (u_s^1)^{-1}$ where the $u_s^p$ with $p = 1,2$
are $2k \times 2k$ graded matrices of the coset $U(k/k)/[U(1)]^{2k}$,
and the $\lambda_s$ and $\overline{\lambda}_s$ are diagonal $2k \times
2k$ matrices related by symmetry. Then $\mbox{trg}(Q \tau_3) = 4
\sum_s (-)^s \mbox{trg} (\lambda_s \overline{\lambda}_s) $ depends
only upon the ``eigenvalues'' $\lambda_s$ and $\overline{\lambda}_s$
of $t_s^{12}$ and $t_s^{21}$, respectively. We write the eigenvalues
in the form $\lambda_{sj}^{\alpha} = i \sin (\theta_{sj}^{\alpha} /2)
\exp ( i \phi_{sj}^{\alpha} )$ and $\overline{\lambda}_{sj}^{\alpha} =
i \sin (\theta_{sj}^{\alpha} /2) \exp ( - i \phi_{sj}^{\alpha})$. The
angle $\theta_{sj}^{\alpha}$ is positive imaginary (positive real) for
$\alpha = b$ ($\alpha = f$, respectively). The transformation is made
unique by the requirement that for fixed $s$ and $\alpha$, the
absolute values $|\theta_{sj}^{\alpha}|$ decrease monotonically with
increasing $j$. We write the matrices $u_s^p$ as products of two
factors obtained by exponentiating the coset generators diagonal
(nondiagonal) in $j$. For the generators diagonal in $j$, this yields
the block--diagonal matrix $w_s^p = \mbox{diag}(w_{sj}^p)$, with
$w_{sj}^p$ belonging to the coset $\mbox{U}(1/1)/[\mbox{U}(1)]^2$ and
parametrized by two anticommuting variables $\gamma_{sj}^p,
\gamma_{sj}^{p*}$. By construction, the variables
$\theta_{sj}^{\alpha}, \phi_{sj}^{\alpha}, \gamma_{sj}^p$, and
$\gamma_{sj}^{p*}$ appear in $T_C$ only in terms of the matrices 
$\chi_{sj} = w_{sj}^1 \lambda_{sj} (w_{sj}^2)^{-1}$, where
$\lambda_{sj}=\mbox{diag}(\lambda_{sj}^{\alpha})$, and in terms of the
symmetry--related quantities $\overline{\chi}_{sj}$.

The Berezinian of the coordinate transformation to polar coordinates
is singular. This is due to the occurrence of the factor $\prod_{s =
  3,4}\prod_j (\cos \theta_{sj}^b - \cos \theta_{sj}^f)^{-2}$. 
Obviously, with the assumed ordering of $|\theta_{sj}^{\alpha}|$,  
singularities occur when at least one of the two pairs $\theta_{sk}^b,
\theta_{sk}^f$, $s=3,4$, vanishes. This is what happens at
$\mbox{trg}(Q\tau_3)=0$. The explicit form of the transformation
$t_s^{12} = u_s^1 \lambda_s (u_s^2)^{-1}$ implies that the integral of
$S(Q)$ over the matrices $u_s^p$ vanishes at the singularities, and is
a regular function of the angles $\theta_{sj}^{\alpha}$ in their
neighborhood. We handle the singularities in analogy to
Refs.~\cite{gos98,plu99}. Lack of space forces us to sketch the
essentials only, see Ref.~\cite{plu00}.

We consider an integral ${\cal I}(kk)$ of the form of Eq.~(\ref{eq1})
but with an infinitesimal neighborhood of size $\eta \rightarrow 0$ of
$\theta_{sk}^b=\theta_{sk}^f=0$, $s=3,4$ removed from the domain of
integration. The removal is accomplished by inserting the Heaviside
functions $\prod_s \theta( | \sin ( \theta_{sk}^b /2) |^2 + | \sin (
\theta_{sk}^f /2) |^2 - \eta )$ in the integrand. The resulting
integral is evaluated in two ways. First, we show that the limit $\eta
\rightarrow 0$ exists. We do so by fixing the order of integrations in
the integral: We integrate over the angles $\theta_{sj}^{\alpha}$ {\it
  after} having integrated over all other variables. The result
differs from the integral ${\cal I} = \overline{ F(E_1) F(E_2) }$ in
Eq.~(\ref{eq1}), however. This is because when expressed in the
original variables, the angles $\theta_{sk}^{\alpha}$ possess
nilpotent parts. Second, we evaluate the same integral by passing from
the coordinates $ z_{sj} = ( \theta_{sj}^{\alpha}, \phi_{sj}^{\alpha},
\gamma_{sj}^p, \gamma_{sj}^{p*} ) $ to the coordinates $ x_{sj} = (
\chi_{sj}^{\alpha  \alpha'}, (\chi_{sj}^{\alpha \alpha'})^* )$, where
$\chi_{sj}^{\alpha \alpha'}$ are elements of the matrices $\chi_{sj} =
w_{sj}^1 \lambda_{sj} (w_{sj}^2)^{-1}$ introduced above. Since the
Berezinian of this transformation removes the singularity from the 
integration measure, the limit can be taken by setting $\eta$ equal to
zero. Expanding the Heaviside functions in powers of the new
anticommuting variables yields ${\cal I}(kk)$ as a sum of two terms. 
The first term is equal to the integral ${\cal I}$. The second term
contains delta functions of $x_{sk}$ in the integrand. These delta
functions are generated by the derivatives of the Heaviside functions. 
Thus, the second term is a ``boundary'' term where some of $x_{sk}$
are put equal to zero in the integrand and removed from the
integration. We conclude that the integral ${\cal I}$ can be found by
evaluating the contributions stemming from this boundary term and from
the ``volume'' term ${\cal I}(kk)$. Evaluating the boundary term in
polar coordinates, we encounter again a singularity. We handle this
singularity in the same fashion as before. Continuing this procedure
finally yields the integral $\overline{ F(E_1) F(E_2) }$ as a sum of
integrals over the polar coordinates generated in the described
way. The sum contains the volume term and a number of boundary terms
(Efetov--Wegner terms) of decreasing complexity. The result for the
autocorrelation function $C_{\cal E}$ reads
\begin{equation}
C_{\cal E} = \sum_{l_3,l_4=0}^{{}\quad \ k \quad \prime}{\cal I}(l_3
l_4) \ .
\label{newCE}
\end{equation}
The prime on the summation sign indicates that the term with $l_3 =
l_4=0$ is omitted. This is because it is cancelled by the unlinked
term $\overline{F(E_1)} \ \overline{F(E_2)}$. The symbols ${\cal
  I}(l_3 l_4)$ stand for integrals of the form of Eq.~(\ref{eq1}),
written in polar coordinates, with all variables $z_{s j}$ carrying
the indices $j > l_s$ put equal to zero in the integrand, and
removed from the integration. The integration over the angles
$\theta_{sj}^{\alpha}$ must be done {\it after} all other integrals
have been worked out. Thus,
\begin{eqnarray}
&&{\cal I}(l_3 l_4)
= \int \mbox{d}\mu_{l_3 l_4}(\theta) S_{l_3 l_4}(\theta) \nonumber \\
&& \qquad \qquad \times \, \exp \bigl( 4 i\pi(\varepsilon/d)
\sum_{sj} (-)^s \mbox{trg} \sin^2 (\theta_{sj}/2) \bigr) \ , 
\label{newI34}
\end{eqnarray}
where $\mbox{d}\mu_{l_3 l_4}(\theta)$ denotes the measure for
integration over the angles $\theta_{sj}^{\alpha}$, and $ S_{l_3
  l_4}(\theta) $ stands for the integral of $S(Q)$ over all remaining
variables. By construction, $S_{l_3 l_4}$ is an even function of each
$\theta_{sj}^{\alpha}$, and, for fixed $s$ and $\alpha$, a symmetric
function of the $\theta_{sj}^{\alpha}$'s with $j \le l_s$. For
$\theta_{s l_s}^b,\theta_{s l_s}^f \rightarrow 0$ at fixed values of
the remaining angles $\theta_{sj}^{\alpha}$, $S_{l_3 l_4}$ tends to
zero with the first power of $ |\theta_{s l_s}^b|^2 + |\theta_{s
  l_s}^f|^2 $.

The entire $\varepsilon$--dependence of the integrals $ {\cal I}(l_3
l_4) $ resides in the exponential in Eq.~(\ref{newI34}). Therefore,
the asymptotic behavior $ ( (\varepsilon/d) \rightarrow \infty ) $ of
these integrals is determined by the behavior of the integrands for
small values of the $\theta_{sj}^{\alpha}$'s. We introduce the
rescaled angles $\tilde \theta_{sj}^{\alpha} = (\varepsilon/d)^{1/2}
\theta_{sj}^{\alpha}$ and expand all terms in the integrand in powers
of $ (\varepsilon/d)^{-1/2} $. Since all terms are even functions of
each $\theta_{sj}^{\alpha}$, the expansions proceed in powers of
$(\varepsilon/d)^{-1}$. As usual, we extend the domain of integration
over $\theta_{sj}^{f}$ to infinity. The asymptotic behavior of ${\cal
  I}(l_3 l_4)$ is given in terms of the leading terms in the
expansions. For the measure and the exponential, these terms can be
straightforwardly found. The detailed form of the leading term in the
expansion of $S_{l_3 l_4}(\theta)$ is specific for the observable
under consideration, and will change when another such function is
considered. However, the general properties of $S_{l_3 l_4}$ discussed
above imply a lower bound on the degree $n_{l_3 l_4}$ of this term. 
For $l_3,l_4 \ne 0$, we find that $n_{l_3 l_4}$ must be larger than or
equal to $2(l_3 + l_4) - 2$; for $l_3 = 0$ ($l_4 = 0$), the
corresponding bound is $(2l_4 - 1)$ ( $(2 l_3 - 1)$, respectively ). 
The asymptotic behavior of the autocorrelation function $C_{\cal E}$
is, therefore, determined by the terms ${\cal I}(l_3 l_4)$ with the
lowest values of $l_3,l_4$, $(l_3,l_4) =  (1,0) (0,1) (1,1)$. The
function $C_{\cal E}$ is symmetric in $E_1, E_2$, and only even powers
of $(\varepsilon/d)^{-1}$ can, therefore, occur in the asymptotic
series. We conclude that the autocorrelation function $C_{\cal E}$
vanishes asymptotically at least as the second inverse power of
$(\varepsilon/d)$, 
\begin{equation}
C_{\cal E} \sim \mbox{const} \times (\varepsilon/d)^{-2} \ .
\label{newCEas}
\end{equation}
This behavior is generic and guarantees ergodicity.

We point out that nowhere in this proof have we used specific
properties of the function $S$ appearing under the integral in
Eq.~(\ref{eq1}). Thus, our argument applies equally to all $4k$--point
functions. But $k$ is an arbitrary positive integer. Hence, our result
applies to {\it all} GUE observables which depend on energy. We have,
however, used that $F(E,H)$ is an observable, hence real, and that
$C_{\cal E}$ is a {\it symmetric} function of $E_1,E_2$. For a non--real
quantity $G(E,H)$, the associated correlation function
$\tilde{C}_{\cal E} = \overline{G(E_1) G(E_2)^*} - \overline{G(E_1)} \
\overline{G(E_2)^*}$ is not symmetric in $E_1,E_2$ and generically
decays asymptotically only like $(\varepsilon/d)^{-1}$. The
$S$--matrix autocorrelation function worked out in Ref.~\cite{ver85}
serves as an example.

We turn to the second case and consider an observable $F(B) = F(H(B))$
which depends on the magnetic field strength $B$. The
magnetoconductance autocorrelation function is an example. The running
average is performed over $B$. The arguments leading to
Eq.~(\ref{erg2}) apply analogously. Ergodicity now requires vanishing
of the autocorrelation function of $F(B)$ at large field differences. 
A novel problem arises, however, in the modeling of the dependence of
the random--matrix ensemble on $B$. For the case of the energy $E$,
there is no such problem because the energy dependence of any
observable arises from its dependence on a retarded or advanced Green
function $(E^{\pm} - H)^{-1}$. This leads to the
$\varepsilon$--dependence displayed in Eq.~(\ref{eq1}) and used 
throughout. In the case of a magnetic field, the Hamiltonian itself
carries the parametric $B$--dependence. We again consider $F(B)$ to be
a $k$--fold correlation function and ask: How do we model the
random--matrix ensembles that describe the Hamiltonians referring to
$2k$ different magnetic field strengths $B \pm \Delta B + \beta_j$, $j
= 1, \ldots, k$? Here, $B$ is some mean magnetic field strength, the
variables $B \pm \Delta B$ correspond to the variables $E_1$ and $E_2$
in the case of an energy--dependent observable, and the variables
$\beta_j$ are analogous to the $\omega_j$'s. We consider the case
where the autocorrelation function is invariant with regard to 
reversing the sign of each of the $\beta_j$'s, and model the $2k$
Hamiltonians by the matrices
\begin{equation}
\label{eq4}
H_{\mu \nu}^{(r,j)} = H_{\mu \nu}^{(I)} + (-)^{r + 1} b_0 \sqrt{1/N}
H_{\mu \nu}^{(II)} + b_j \sqrt{1/N} H_{\mu \nu}^{(III)} \ .
\end{equation}
Here, $b_0$ is proportional to $\Delta B$, and the $b_j$'s with $j =
1, \ldots, k$ are proportional to the $\beta_j$'s, respectively. The
index $r$ takes the values $r = 1,2$ for $B + \Delta B$ and $B -
\Delta B$, respectively. The three GUE ensembles $H^{(l)}$ with $l =
I,II,III$ are taken to be uncorrelated. Eq.~(\ref{eq4}) is a 
straightforward generalization of the usual model applied in the case
of two different magnetic field strengths, see Ref.~\cite{gos98}. 

We consider the limit $|\Delta B| \rightarrow \infty$ of the
autocorrelation function $ C_{\cal B} $ $ = \overline{ F(B + \Delta B)
  F(B - \Delta B) } $ $ - \overline{ F(B + \Delta B) } \ \overline{
  F(B - \Delta B) } $, keeping the $\beta_j$'s fixed. To calculate $
C_{\cal B} $, we again use the supersymmetry technique and arrive at
an expression which has the form of Eq.~(\ref{eq1}) except that the
exponential function under the integral is replaced by $\exp ( -
[b_0^2/2] \ {\rm trg} (Q \tau_3)^2 )$. Moreover, the function $S$ now
depends on the field variables $b_j, j = 1, \ldots, k$ rather than on
the energies $\omega_j$. We proceed as before and arrive at an
equation which has the form of Eq.~(\ref{newCE}) except that now the
integrals ${\cal I}(l_3,l_4)$ contain the exponential $\exp ( 2 b_0^2
\sum_{sj} {\rm trg} \ \sin^2 \theta_{sj} )$. This expression is a
symmetric function of $\Delta B$ and, thus, automatically obeys the
symmetry property of $ C_{\cal B} $. Proceeding as before we finally
conclude that the leading term in the asymptotic series for $C_{\cal
  B}$ stems generically from the Efetov-Wegner terms ${\cal I}(10)$
and ${\cal I}(01)$ and is of order $b_0^{-2}$. This implies that for
$|b_0| \rightarrow \infty$, we get
\begin{equation}
\label{eq5}
C_{\cal B} \sim \mbox{const} \times b_0^{-2} \ .
\end{equation} 
This asymptotic behavior is sufficient to guarantee ergodicity.

In summary, we have shown ergodicity to hold for a wide class of
ensembles of unitary random matrices and for all observables which
depend either on energy or magnetic field strength. We expect that our
proof applies likewise to the orthogonal and to the symplectic
ensembles.

Thanks are due T. Guhr who informed us in the course of this work that
several years ago, he designed another proof of ergodicity also based
on supersymmetry, and who helped us with informative discussions. Z.P.
thanks the members of the Max-Planck-Institut f\"ur Kernphysik in
Heidelberg for their hospitality and support, and acknowledges support
by the grant agencies GACR and GAUK in Prague.

\end{document}